\documentclass[pra,reprint,superscriptaddress]{revtex4-1}
\usepackage{amssymb}
\usepackage[latin1]{inputenc}
\usepackage{amsmath}
\usepackage{braket}
\usepackage{amsfonts}
\usepackage[margin=1.0in]{geometry}
\usepackage{graphicx}

\begin{document}

\title{Stability thresholds and calculation techniques for fast entangling gates on trapped ions}
\date{\today}

\author{C D B Bentley}
\email{christopher.bentley@anu.edu.au}
\affiliation{Department of Quantum Science, Research School of Physics and Engineering, The Australian National University, Canberra, Australia}

\author{R L Taylor}
\affiliation{Department of Quantum Science, Research School of Physics and Engineering, The Australian National University, Canberra, Australia}
\author{A R R Carvalho}
\affiliation{Department of Quantum Science, Research School of Physics and Engineering, The Australian National University, Canberra, Australia}
\affiliation{ARC Centre for Quantum Computation and Communication Technology, The Australian National University, Canberra, Australia}
\author{J J Hope}
\affiliation{Department of Quantum Science, Research School of Physics and Engineering, The Australian National University, Canberra, Australia}

\begin{abstract}

Fast entangling gates have been proposed for trapped ions that are orders of magnitude faster than current implementations.
We present here a detailed analysis of the challenges involved in performing a successful fast gate.
We show that the RWA is a stable approximation with respect to pulse numbers: the timescale on which we can neglect terms rotating at the atomic frequency is negligibly affected by the number of pulses in the fast gate. 
In contrast, we show that the laser pulse instability does give rise to a pulse-number dependent effect; the fast gate infidelity is compounded with the number of applied imperfect pulses.
Using a dimensional reduction method presented here, we find bounds on the pulse stability required to achieve two-qubit gate fidelity thresholds.

\end{abstract}

\maketitle

\section{Introduction}

A two-qubit entangling gate is an essential component of any quantum information processing (QIP) system~\cite{DiV00FP}. 
Fast gates for trapped ions using controlled large momentum kicks offer a significantly faster operation timescale than traditional gates requiring spectral resolution of sidebands~\cite{GZC03PRL, Duan04PRL, GZC05PRA, Bent13NJP, Bent15NJP, Mizr13APB}.
This in turn leads to simpler gate adaption for long ion crystals or more complex geometries~\cite{Duan04PRL,Zhu06PRL,Lin09EPL,Zou10PLA}, with relatively invariant schemes required for sufficiently fast gates~\cite{Zhu06EL,Bent15NJP}.
There has been recent progress towards the implementation of pulsed fast gates in the production of the required high repetition-rate pulsed lasers~\cite{Petr14OE} and their application to perform a single-qubit gate~\cite{Camp10PRL}, as well as spin-motion entanglement~\cite{Mizr13PRL}.
In this paper we outline challenges for performing a complete fast gate protocol, and present both the techniques for quantifying gate fidelity subject to imperfections, as well as the required thresholds in laser stability and pulse times to perform high-fidelity gates.

For implementation of a fast gate, and certainly for considering their application to large-scale algorithms, detailed analysis of the stability requirements for the trap and control are critical.
Error correction can be applied between gate operations, however individual gate operations are required to meet high-fidelity thresholds~\cite{NC00,Knill05N,Benh08NP}.
The scheme proposed by Garc\'ia-Ripoll, Zoller and Cirac (GZC scheme)~\cite{GZC03PRL} and the Fast Robust Antisymmetric Gate (FRAG scheme)~\cite{Bent13NJP} were shown to have very high fidelity and robustness in~\cite{Bent15NJP}, and we focus on these schemes for our error analysis.
While perfect GZC and FRAG gates are independent of the initial motional states, errors in the gate are enhanced according to the mean vibrational mode occupation, as shown in~\cite{GZC05PRA,Bent15NJP}.
Certain gate error sources have been considered: the effects of trap anharmonicity on both schemes~\cite{GZC03PRL,GZC05PRA,Bent15NJP}, dissipation effects on the GZC scheme~\cite{GZC05PRA} and laser control errors in the FRAG scheme such as insufficient laser repetition rates and pulse timing or direction errors~\cite{Bent13NJP}.
Only a preliminary analysis of pulsed fast gate errors due to pulse area imperfections has been performed, despite the conclusion that such errors are significant~\cite{GZC03PRL,Bent13NJP}. 
Furthermore, a phase stability analysis is still required.

Laser phase dependence in each momentum kick comprising the fast gates arises when the rotating wave approximation (RWA) is no longer valid due to short pulse durations relative to the atomic transition frequency.
This leads to imperfect population transfer between internal states.
Short pulse durations are necessary such that the total motional evolution during the pulses is negligible. 
Thus the combined duration from all of the applied pulses is required to be much shorter than the trap period, which is on the order of 1~$\mu$s. 
Few pulses or very short pulse durations seem preferable, however increasing the number of pulse pairs in a fast gate improves the gate speed, fidelity and scaling with the number of ions.
The RWA provides the lower bound for pulse durations for fast gates, and examining the dependence of this lower bound on the number of pulses in a gate is essential for applications of fast gates to QIP. 

Significant infidelity also arises from imperfect applied pulses.
Ideal pulses keep the internal qubit states invariant throughout the phase gate and restore the initial motional state at the end of the gate, however imperfect pulses cause internal state transfers as well as occupation of a range of motional levels after the gate.
Random errors in the pulse duration were considered in~\cite{GZC03PRL} for a four-kick sequence, and in~\cite{Bent13NJP} a worst-case error bound was calculated using perturbation theory for small errors in the pulse area and low numbers of pulse pairs.  
The perturbation technique was used for just four pulse pairs, and fails well before 100 pulse pairs with error on the order of 1\% in the pulse area~\cite{Bent13NJP}.
It was concluded that imperfect pulse area will limit the fidelity of fast gates; a more complete analysis of the required pulse stability is necessary for gate implementation.

It is possible to model the full dynamics of a gate without the RWA or with pulse area imperfections, and thereby directly calculate the gate fidelity.
However, the Hamiltonian operator required for this calculation has a dimension given by the square of the full state vector dimension, which includes both the internal qubit states and the vibrational mode states for each shared mode.
In the ideal-pulse case, the complexity is vastly reduced by simplifying the requirements for performing a high-fidelity gate to three control conditions~\cite{GZC03PRL}.
In this paper, we present a simplification method for imperfect gates that permits fidelity calculation for large momentum kicks in traps with many ions and a corresponding number of shared motional modes.
Our method is presented in Section~\ref{sec:method} following a review of fast gates.
In Section~\ref{sec:nrwa}, we apply this fidelity calculation technique to explore the effect of pulse number on the phase dependence of the gate with short pulse durations.
This provides a minimum pulse duration bound for high-fidelity fast gates composed of varying numbers of pulses.
In Section~\ref{sec:pulsearea}, we apply our method to consider imperfect pulse areas comprehensively.  We introduce the errors in the atom-light evolution unitaries, and construct the imperfect gate evolution operators to directly compute the fidelity.
This gives us an accurate measure of fidelity for large numbers of pulses.
Finally, we present our conclusions in Section~\ref{sec:conc}.

\section{Quantifying fast gate errors} \label{sec:method} 

We present the fast gate mechanism and summarise the GZC and FRAG gate schemes, followed by a general fidelity calculation method for two trapped ions as well as a two ion gate in a longer ion crystal.

\subsection{Gate dynamics and limitations}

\begin{figure*}[t] 
     \centerline{\includegraphics[width= 2\columnwidth]{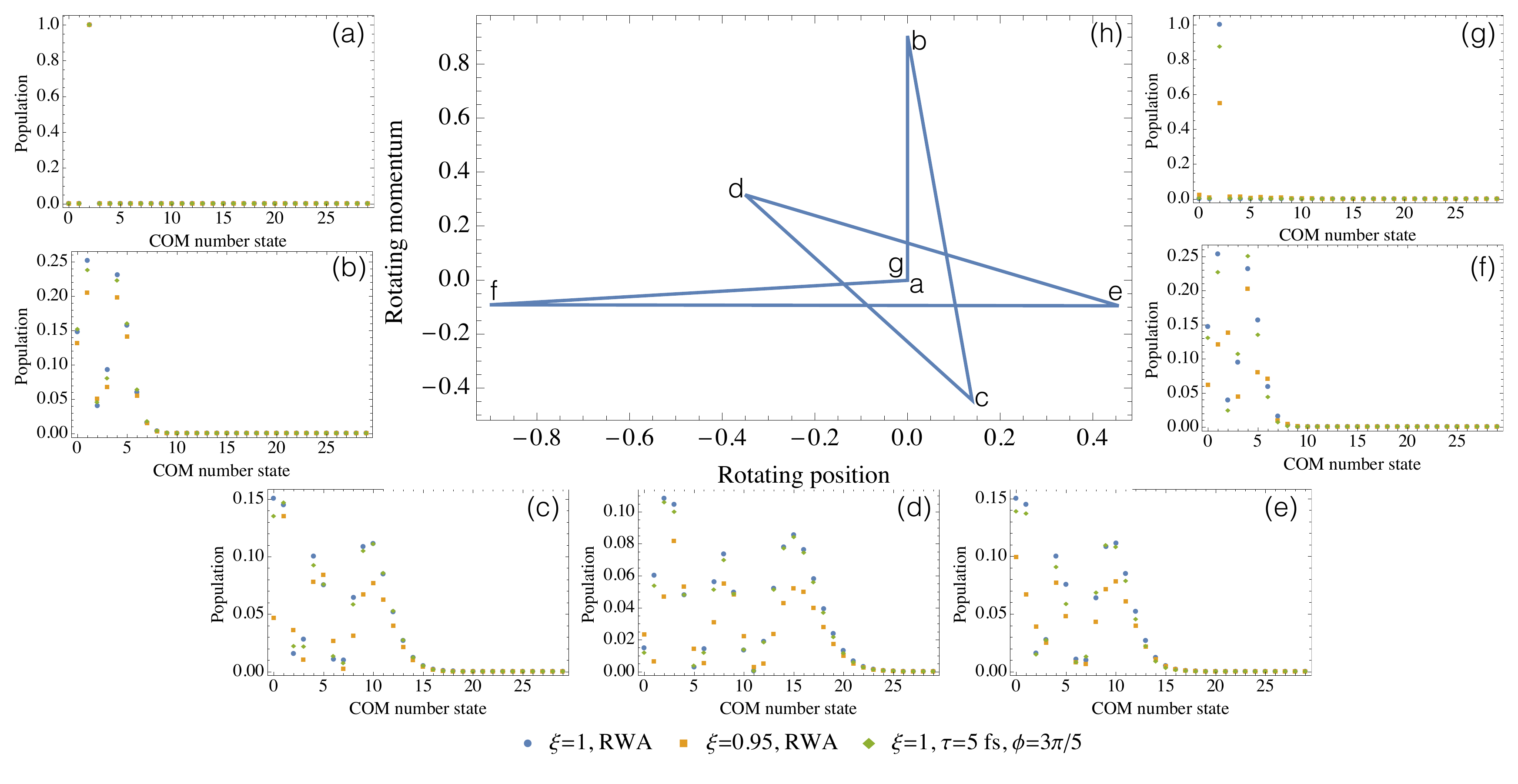}}
   \caption[Errors in the fast gate evolution dynamics]{
   (h): Centre-of-mass (COM) phase-space trajectory for the centre of a coherent state during the gate operation.
   The sides of the trajectory correspond to momentum kicks.
   The angle of each vertex corresponds to free evolution between kicks, and marks the gate evolution point for the other subfigures, from a$\rightarrow$g.
   (a)$\rightarrow$(g): Population occupying COM mode number states, for both ions in the excited state, at different points during the GZC gate operation, with $n=1$ (14 total pulse pairs).
   Blue circles represent an ideal gate, satisfying the rotating wave approximation (RWA) and with no pulse area imperfections given by $(1-\xi)$.
   A gate with systematic pulse imperfections ($\xi=0.95$) is also considered (yellow squares), as well as a gate where the RWA is invalid (green diamonds) such that the pulse duration and laser phase must be defined ($\tau = 5$~fs, $\phi = 3 \pi /5$).
   After the non-ideal gates, population is lost to other internal states, and some of the population is imperfectly restored to the initial COM state, $|2\rangle_c$, the second excited number state.
   Values were chosen to illustrate the effects of these errors.
      The point in the gate operation described by subfigures (a)$\rightarrow$(g) corresponds to a$\rightarrow$g marked in subfigure (h).
   }
   \label{fig:gzcmodeevol}
\end{figure*}

Fast gates operate in the strong-coupling regime, where the laser coupling is much greater than the trap frequency, $\Omega \gg \nu$.
In this regime, multiple number states of each shared motional mode are excited by pairs of counter-propagating laser pulses, as shown in Figure~\ref{fig:gzcmodeevol}.
These $\pi$-pulse pairs provide momentum kicks such that a closed trajectory in phase-space is described for the centre of a coherent state, as in Figure~\ref{fig:gzcmodeevol}~(h).
The area enclosed in each mode's phase-space determines a conditional phase applied to different two-qubit computational states.

The evolution of an ideal fast gate can be described as alternating displacement and rotation operators in phase space for each motional mode.
A displacement operator for mode $p$ is described by
\begin{align}
\hat{D}_p(\alpha) = \exp [ \alpha a_p^\dagger - \alpha^* a_p ],
\end{align}
for a displacement of $\alpha$, where $a_p$ is the mode annihilation operator.
Under the RWA, pairs of counter-propagating $\pi$-pulses give rise to mode displacement operators as follows~\cite{Bent15NJP}:
\begin{align}
U_{\text{kick}} &= e^{-2 i z k (x_1 \sigma_1^z + x_2 \sigma_2^z)} \label{eq:ukick1} \\
&= \Pi_{p=1}^L \hat{D}_p (-2 i z (b_1^{(p)} \sigma_1^z + b_2^{(p)} \sigma_2^z) \eta_p ), \label{eq:ukick2}
\end{align}
when there is negligible motional evolution between the two pulses.
Here $z$ is the direction of the first pulse in the pair, $k$ is the laser wavenumber, $x_i$ is the position operator for ion $i$ and $\sigma_i^z$ is the usual Pauli $Z$ operator acting on ion $i$.
There are $L$ motional modes corresponding to $L$ ions in the crystal, and $b_i^{(p)}$ is the ion-mode coupling coefficient between ion $i$ and mode $p$.
The Lamb-Dicke parameter $\eta_p$ for mode $p$ is given by
\begin{align}
\eta_p = k \sqrt{\frac{\hbar}{2 M \nu_p}},
\end{align} 
for ion mass $M$ and mode frequency $\nu_p$.

Two main causes of imperfect momentum kicks to the ions come from counter-propagating pulses applied with area not equal to $\pi$, or from breaking the RWA through short pulse durations.  
These imperfections are shown in Figure~\ref{fig:gzcmodeevol}, which illustrates their effect at each stage in the fast gate evolution process.

The free motion of the ions, and their motional modes, corresponds to rotation operators in each mode $p$'s phase space:
\begin{align}
U_{p,mot} = e^{-i \nu_p \delta t_k a_p^\dagger a_p}, \label{eq:freemotun}
\end{align} 
where $\delta t_k$ is the time between the $k$th and $(k+1)$th momentum kicks.

The displacements and free rotations are determined according to particular pulse schemes.
These schemes satisfy the required gate conditions: (1) conditional phase evolution according to the two-qubit gate described by
\begin{align}
U_{\text{gate}} = e^{i \frac{\pi}{4} \sigma_1^z \sigma_2^z} \label{eqn:idun}
\end{align}
for a gate applied to ions 1 and 2, and (2) no motional dependence, such that the initial motional state is restored following the gate operation.
The GZC and FRAG schemes are characterized by pulse pairs $\underbar{z}$ applied at times $\underbar{t}$, interspersed with free evolution.
For the FRAG scheme~\cite{Bent13NJP}:

\begin{center}
\begin{tabular}{ccccccc}
$\underbar{z}$ &= ($-n$, &$2n$, &$-2n$, &$2n$, &$-2n$, &$n$) \\
$\underbar{t}$ &= ($-\tau_1$, &$-\tau_2$, &$-\tau_3$, &$\tau_3$, &$\tau_2$, &$\tau_1$).
\end{tabular} \\
\end{center}

At time $-\tau_1$, $n$ counter-propagating pulse pairs are applied along the trap axis (aligned with the $z$ axis) to provide a $2 n \hbar k$ momentum kick in the $-z$ direction.

The GZC scheme~\cite{GZC03PRL,GZC05PRA} is characterised as follows: 

\begin{center}
\begin{tabular}{ccccccc}
$\underbar{z}$ &= ($-2n$,	  &$3n$,  &$-2n$, &$2n$, &$-3n$, &$2n$) \\
$\underbar{t}$ &= ($-\tau_1$,  &$-\tau_2$,  &$-\tau_3$, &$\tau_3$, &$\tau_2$, &$\tau_1$).
\end{tabular} \\
\end{center} 

The integer $n$ determines the gate time $T_G$, which scales optimally with the total number of pulses in the scheme $N_p$ as $T_G \propto N_p^{-2/3}$~\cite{GZC05PRA,Bent15NJP}.

The FRAG and GZC schemes consist of $10n$ and $14n$ pulses respectively. 
The FRAG scheme has a state-averaged fidelity, as defined in~\cite{Bent15NJP}, of 0.96 for $n=1$, and 0.995 for $n=2$, while for higher $n$ the infidelity is below $10^{-8}$.
We neglect the low-fidelity $n=1$ case of the FRAG scheme, which obscures the stability analysis.
The GZC scheme, with higher total numbers of pulses for each $n$, achieves infidelity on the order of $10^{-5}$ for $n=1$, and infidelity below $10^{-8}$ for higher $n$.

The scaling of errors with the number of pulses is examined for both the FRAG and GZC schemes.
We explore the effects of errors on schemes with low pulse-numbers using the $n=1$ GZC scheme due to its high fidelity.
While more robust for lower numbers of pulses, the GZC scheme is slower than the FRAG scheme for $n \geq 2$, as shown in~\cite{Bent13NJP}.
The effects of finite laser repetition rates on these schemes were explored in~\cite{Bent13NJP, Bent15NJP}, where it is shown that for repetition rates of around 300~MHz, even a gate with perfect $\pi$-pulses has non-negligible infidelity.
Faster repetition rates have robust fidelities, particularly for the two-qubit case.
Errors due to imperfect pulse areas or from breaking the RWA affect any scheme regardless of repetition rate; in this paper we consider these errors independently by assuming an infinite repetition rate.
Furthermore, this approximation allows a clear analysis of the relationship between these errors and number of applied pulses.
The methods in this paper can be applied using particular, finite repetition rates to model the errors in an experiment more precisely.

To model the effect of imperfect pulses or an invalid RWA, we expand the appropriate unitary operator for the applied gate, $U_{re}$, in the number basis.
We can then observe the phase-space evolution during the gate process, and calculate the fidelity of the gate.
While coherent states are preserved by the momentum kicks and rotations, the momentum kicks deform an initial number state to spread across many modes.
At the end of a high-fidelity gate, however, this spread resolves back into the initial number state, as shown in Figure~\ref{fig:gzcmodeevol}(a)-(g).

\subsection{Fidelity calculation: dimensional reduction}

To assess the impact of particular errors, we use the state-averaged fidelity as a measure of the gate performance.
The fidelity of a pure state $|\psi \rangle$ with respect to a density matrix $\sigma$ is given by the state overlap \cite{Gilc05PRA}:
\begin{align}
F = \langle \psi | \sigma | \psi \rangle.
\end{align}
For an initial state $|\phi_i\rangle$, the final state following the ideal gate operation $U_{id}$ is given by
\begin{align}
|\psi \rangle &= U_{id} |\phi_i \rangle,
\end{align}
and the final density matrix following the real, imperfect, operation $U_{re}$ is given by
\begin{align}
\sigma &= U_{re} |\phi_i \rangle \langle \phi_i | U_{re}^\dagger.
\end{align}
The state-averaged fidelity is thus
\begin{align}
F = \int_{\phi_i} | \langle \phi_i | U_{id}^\dagger U_{re} | \phi_i \rangle |^2,
\end{align}
integrating uniformly over the unit hypersphere described by the initial state with arbitrary coefficients $a_{jk}$:
\begin{multline} 
|\phi_i \rangle = \left( a_{00} |gg\rangle + a_{01} |ge\rangle + a_{10} |eg\rangle + a_{11} |ee\rangle  \right) \\
\otimes |n_c n_r \rangle.
\end{multline}
The initial motional state is the number product state $|n_c\rangle \otimes | n_r \rangle$ for the centre-of-mass (COM) and stretch modes respectively.
The motional inner product is stricter than the computational fidelity of~\cite{Bent15NJP}, with a stronger motional restoration requirement that population must be restored to the initial number state for each mode at the end of the gate operation.
This is a convenient choice for our number basis, and directly considers effective heating caused by the gate to be infidelity.

The ideal gate operation of equation~(\ref{eqn:idun}), with duration $T_G$, applies a state-dependent phase while preserving the internal and motional states: 
\begin{multline}
U_{id} |\phi_i\rangle = \left( e^{i \pi/4} a_{00} |gg\rangle + e^{-i \pi/4} a_{01} |ge\rangle \right.  \\
  + \left. e^{-i \pi/4} a_{10} |eg\rangle + e^{i \pi/4} a_{11} |ee\rangle  \right)   \\
  \otimes e^{-i T_G (\nu_c n_c + \nu_r n_r)} |n_c n_r \rangle, \label{eq:idact}
\end{multline}
where the motional component is global phase corresponding to free evolution for each mode.

The real gate is a more complex operation on both the computational and motional states, and we consider the error to first order in the small gate imperfection.
Since the ideal gate does not transform the basis states but just applies a phase, a real gate approximating the ideal operation has only small population transfer between internal states.
The gate schemes are designed to restore the motional states for preserved internal states; only a fraction of the motional state population (to second order in the error) will be restored for altered internal states with changed state-dependent displacement operators.
These terms with changed internal states thus provide a second-order correction to the fidelity, which we neglect here.

Similarly, an ideal counter-propagating pair of pulses acting on two ions with the same internal state affects only the COM mode.
An imperfect pair of pulses may alter the stretch mode to some small degree; a first-order error term.
This perturbation to the stretch mode also has only a small effect on the fidelity; only a fraction of the perturbation is expected to be restored to its initial motional state and this second-order contribution is neglected.
Ions with opposite internal states are assumed to have an invariant COM mode, with the gate acting on the stretch mode.

For the basis state 
\begin{align}
a_{00} |g g\rangle \otimes | n_c n_r \rangle \equiv a_{00} |g g n_c n_r \rangle,
\end{align}
the fidelity inner product element is thus
\begin{multline}
|a_{00}|^2 \langle g g n_c | U_{id}^\dagger U_{re} | g g n_c \rangle  \\
  = |a_{00}|^2 e^{-i \pi/4} \langle g g n_c | U_{re} |g g n_c \rangle,
\end{multline}
where the stretch mode is allowed to evolve freely by the ideal and real unitaries, and thus cancels from the inner product.
Only the population retained in the computational ground state of both ions is retained in the fidelity term.
The unitaries act symmetrically on $|e e \rangle$, and the same symmetry between $|e g\rangle$ and $|g e \rangle$ allows us to simplify our full fidelity expression:
\begin{multline} 
F = \int_{\phi_i} \left| (|a_{00}|^2 + |a_{11}|^2) e^{i \nu_c T_G n_c -i \pi/4} \right. \\
\times \langle g g n_c | U_{re} | g g n_c \rangle + (|a_{01}|^2 + |a_{10}|^2) e^{i \nu_r T_G n_r + i \pi/4} \\
  \left. \times  \langle g e n_r | U_{re} | g e n_r \rangle   \right| ^2. \label{eq:fid1}
\end{multline}
For each internal state of the two qubits, the effect of the real gate on just a single motional mode contributes to the fidelity expression.
We thus expand $U_{re}$ for each mode independently, and accordingly cancel the motional phase term in the ideal unitary from free evolution of the other mode.

A general position operator decomposition for two trapped ions is described by
\begin{align}
k x_i = b_i^{(c)} \eta_c (a_c + a_c^\dagger) + b_i^{(r)} \eta_r (a_r + a_r^\dagger), \label{eq:posdecomp}
\end{align}
where the subscript $c$ describes the COM mode, and subscript $r$ describes the stretch mode.
The coupling operators for two trapped ions are
\begin{align}
b^{(c)} &= (\frac{1}{\sqrt{2}},\frac{1}{\sqrt{2}}) \\
b^{(r)} &= (-\frac{1}{\sqrt{2}},\frac{1}{\sqrt{2}}),
\end{align}
with the $j$th vector element representing the coupling for ion $j$, and mode frequencies of $\nu$ and $\sqrt{3} \nu$ for the COM and stretch modes respectively.

The real gate unitary $U_{re}$ is applied to both modes, however to consider first-order errors we can treat it as separable for each mode.
This allows us to apply a single-mode expansion of $k x_i$ for each internal state of the qubits.
For the states $|gg\rangle$ and $|ee\rangle$, the terms in $U_{re}$ contributing to the fidelity equation~(\ref{eq:fid1}) are given by
\begin{align}
k x_i = b_i^{(c)} \eta_c (a_c + a_c^\dagger),
\end{align} 
while for $|ge\rangle$ and $|eg \rangle$,
\begin{align}
k x_i = b_i^{(r)} \eta_r (a_r + a_r^\dagger).
\end{align}

Our separable representation of $U_{re}$ reduces the dimension of the state vector by a factor given by the number of required motional basis states.
The Hamiltonian is reduced in dimension by the square of this factor.
We use the number basis to model the state evolution, and truncate the basis such that negligible population occupies the maximal basis states during the gate operation.
This truncation occurs for higher phonon numbers with larger numbers of applied pulses in a gate corresponding to larger momentum kicks.
We truncate our number basis at 50 states for $n=1$ for each gate, 70 states for $n=2$ and $n=5$, and 130 states for $n=10$.
The single-mode analysis thus reduces the dimensions for calculating the state vector evolution by a factor of around $100$, depending on the number of applied pulses.
The state vector's dimension is four times the dimension of the number basis, due to the two basis internal states for each ion.

\subsection{Extending fidelity calculation to larger traps}

Performing fast gates within a large quantum processor requires analysis of gate imperfections in traps with larger numbers of ions.
The number of motional modes is equal to the number of trapped ions.
The position decomposition can be generalised from equation~(\ref{eq:posdecomp}) for arbitrary numbers of ions, and we again apply the approximation of separable motional modes in a harmonic trap.

The ideal and real unitaries can be written as a separable product of the operation on each mode: 
\begin{align}
U_{id}&=\prod_{p=1}^L U_{id,p}\\
U_{re}&=\prod_{p=1}^L U_{re,p},
\end{align}
where $L$ is the number of modes and
\begin{align}
U_{id,p}&=e^{i\phi_p\sigma_1^z \sigma_2^z}, 
\end{align}
up to global phase.
This simplifies the fidelity $F$ expression to require only the product of the mode-dependent fidelities $F_p$ up to first order in the error term:
\begin{align}
F &\simeq \int_{\phi_i} \left| \Pi_p F_p \right|^2,\\
F_p &= \bra{\phi_i}U^\dagger_{id,p}U_{re,p}\ket{\phi_i}. 
\end{align}

We can calculate the contributions $U_{re,p}$ of each mode to the real gate unitary, using an appropriately truncated single-mode number basis for each calculation. 
For a high-fidelity gate scheme, the ideal phase contribution for each mode is given by~\cite{Bent15NJP} 
\begin{equation}
\phi_p=8\eta_p^2\sigma_1^z \sigma_2^z b_1^{(p)} b_2^{(p)} \sum_{m=2}^N\sum_{k=1}^{m-1}z_m z_k\sin(\nu_p(t_m-t_k)),
\end{equation}
where $z_k$ is the number of pulse pairs applied at time $t_k$ determined by the gate scheme.
For a high-fidelity gate, $\sum_p\phi_p\approx\frac{\pi}{4}$.

Our method to estimate fidelity in this fashion is to first calculate the real unitaries for each mode $U_{re,p}$ and the ideal phases $\phi_p$.
The individual mode expansion of the gate unitary ensures a low-dimensional state vector, as required for computation.

\section{Breaking the RWA} \label{sec:nrwa}

Pulses with duration on the order of the atomic transition period $2 \pi/ \omega_{at}$ render the RWA invalid, and cause infidelity in fast gate schemes which rely on the RWA.
In this section we apply our fidelity calculation method to explore the tradeoff in pulse duration between performing large numbers of pulses in a short time for fast, high-fidelity gates, while staying in the regime where the RWA holds.
Gates significantly faster than the trap evolution period ($\sim 1$~$\mu$s) require large numbers of pulses, which must thus have very short durations.
We demonstrate that the valid RWA regime is altered little by the number of applied pulses in a gate.

We perform fast gates using short pulses of varying duration without performing the RWA to investigate the regime where the approximation holds.
The gate should also be independent of the optical phase $\phi$ \cite{Stea14NJP}, and we quantify the pulse lengths required for phase-independence.
In the interaction frame with respect to the internal states of a single ion, ion $1$ marked by subscripts, the atom-light interaction Hamiltonian is
\begin{multline}
H'_1 = \frac{\hbar \Omega}{2} (\sigma_1^+ e^{-i(kx_1 - (\omega_L+\omega_{\text{at}}) t +\phi)} + \sigma_1^+ e^{i(kx_1 - \delta t +\phi)}  \\
	 + \sigma_1^- e^{-i(kx_1 - \delta t +\phi)} + \sigma_1^- e^{i(kx_1 - (\omega_L+\omega_{\text{at}}) t +\phi)}  ), \label{eq:nrwa}
\end{multline}
where $\delta = (\omega_L - \omega_{\text{at}})$.
Typical atomic frequency transitions are on the order of $\omega_{\text{at}} \sim 2\pi \times 10^{15}$~Hz, and the fast rotating terms can be neglected following the RWA. 
Pulse durations are typically assumed to be much longer than the rotation period, $\tau (\pi \times 10^{15}) \gg 1$.

We focus on resonant transitions where $\delta=0$ for simplicity, such that $\omega_L = \omega_{at}$.
Assuming constant $\Omega$ and a perfect $\pi$-pulse, such that $\Omega = \pi/\tau$, the unitary operator from equation~(\ref{eq:nrwa}) for a single ion is:
\begin{multline} 
U_{\text{pulse},1} = \text{exp} \left[ \frac{-i \pi}{2 \tau} \left( \int_{t_i}^{t_f} \sigma_1^+ e^{-i(k x_1 - 2 \omega_{at} t + \phi)} dt \right. \right. \\
 \left. \left. + \int_{t_i}^{t_f} \sigma_1^- e^{i(k x_1 - 2 \omega_{at} t + \phi)} dt + \tau \sigma_1^+ e^{i(kx_1+\phi)} \right. \right.  \\ 
\left. \left.  + \tau \sigma_1^- e^{-i(kx_1+\phi)}   \right) \right] , \label{eq:unrwa}
\end{multline}
for a pulse of duration $\tau = t_f - t_i$, where
\begin{multline}
\int_{t_i}^{t_f}  \sigma_1^+ e^{-i(kx_1 - 2 \omega_{at} t + \phi)} dt \\
= \frac{-i \sigma_1^+}{2 \omega_{at}} \left( e^{-i(k x_1 - 2 \omega_{at} t_f + \phi)}  - e^{-i(k x_1 - 2 \omega_{at} t_i + \phi)}    \right).
\end{multline}
For two ions, the interaction Hamiltonian is $H'_1+H'_2$ for ions 1 and 2, using $H'_i$ from equation~(\ref{eq:nrwa}).
The pulse unitary operator $U_{\text{pulse},1}$ is similarly extended, and combined with the motional free evolution unitary, equation~(\ref{eq:freemotun}), to construct the real gate unitary $U_{re}$.
This allows us to explore the validity of the RWA for different pulse lengths by solving for the phase dependence and fidelity.

\begin{figure*}[t!] 
     \centerline{\includegraphics[width=2 \columnwidth]{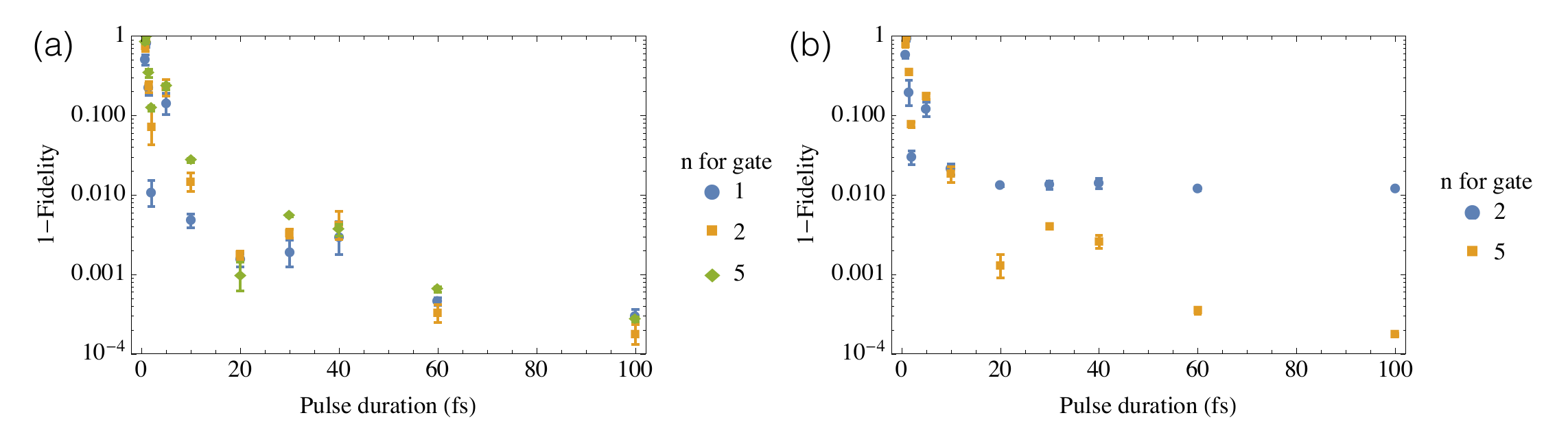}}
   \caption[Breaking the RWA: fast gate fidelities with short pulses]{Infidelity following the (a) GZC and (b) FRAG gate operations with different numbers of pulses, governed by $n$.
   The effect of changing the duration of the pulses composing the gate is shown.
   The initial motional state is $|2\rangle_c |2\rangle_r$, the second excited number state for each mode.
   We determine the mean and the standard deviation (error bars) by varying the phase $\phi$ for a given pulse duration $\tau$.
   }
   \label{fig:fiddurn}
\end{figure*}

Figure~\ref{fig:fiddurn} shows the effect of short pulse duration on gate fidelity.
For short pulse duration, the fidelity decreases as more pulses are applied for the FRAG and GZC gates.
The mean infidelity is plotted for varying phase $\phi$, and error bars mark a standard deviation in infidelity due to phase dependence.
The FRAG gate with $n=2$ has fidelity of 0.988, and approaches this value with a standard deviation less than $10^{-3}$ for pulse lengths $\tau > 40$~fs.
The GZC scheme and the FRAG scheme for $n=5$ have fidelity above 0.999 and standard deviation less than $10^{-4}$ for $\tau \geq 60$~fs.

Pulses much longer than the atomic transition period are accurately described under the RWA, and the number of pulses in the gate does not significantly alter this threshold.
For gates with increasing speed or scalability with the number of ions, large numbers of pulses must be performed much faster than the trap motional frequency, $\nu/(2 \pi) \simeq 1$~MHz, or even much faster than $10$~ns for momentum application schemes exciting short-lived atomic levels~\cite{Gerr08EPJ}. 
This provides five orders of magnitude between a safe pulse duration $\sim 100$~fs and the lifetime of typical short-lived levels, such as $P_{3/2}$ in $^{40}$Ca$^+$.

\section{Imperfect pulses} \label{sec:pulsearea}

Significant errors also arise from imperfect $\pi$-pulses, which construct the momentum kicks fundamental to fast gates.
$\pi$-pulses with arbitrarily high fidelity can be constructed using composite pulses~\cite{Toro11PRA,Ivan11OL}; laser repetition rates must be sufficiently high to accommodate the pulse components in this approach.
In this section, we consider the impact of infidelity in the $\pi$ rotations on the full gate fidelity.
Imperfect $\pi$-pulses cause imperfect state transfer, errant momentum kicks and acquired phase infidelity.

While different methods for performing $\pi$-pulses have varying robustness to laser fluctuations, the pulse rotation fidelity for any method has a fixed relation to the full gate fidelity.
We consider here the simplest case of square pulses to calculate the relation between rotation fidelity and gate fidelity. 
To model the imperfect gate process, we assume a suitable pulse length for the RWA, with $\delta =0$:
\begin{align} 
H_{\text{RWA}}' = \frac{\hbar \Omega}{2} (\sigma_+ e^{i(kx + \phi)} + \sigma_- e^{-i(kx + \phi)}  ).
\end{align}

For $\Omega$ constant in time, a $\pi$-pulse satisfies $\Omega \tau = \pi$, for a pulse duration $\tau$.
An approximate $\pi$-pulse satisfies $\Omega \tau = \xi \pi$, with $\xi \simeq 1$.
The unitary corresponding to the pulse applied to a single ion follows:
\begin{align}
U_{\text{pulse}} 
 = e^{\frac{-i \xi \pi}{2} (\sigma_+ e^{i(kx+\phi)} + \sigma_- e^{-i(kx+\phi)} )}.
\end{align}
Reversing the pulse direction changes the sign of $k$ in the evolution operator.
The pulse rotation fidelity can be found for ideal and real pulse unitaries $U'_{\text{pulse}}$ and $U_{\text{pulse}}$ respectively:
\begin{align}
\text{R. Fid.} &= \text{Min}_{\psi_i} \left| \langle \psi_i | (U'_{\text{pulse}})^\dagger U_{\text{pulse}} | \psi_i \rangle \right|^2 \\
&\simeq 1 - \frac{(1-\xi)^2 \pi^2}{4},
\end{align}
up to third order in $(1-\xi) \pi/2$.

Assuming that the same laser produces each pulse, and that phase drift is minimal during the gate duration ($<1$~$\mu$s), $\phi$ is fixed.
We fix $\xi$ to be constant during a gate operation to find the systematic error effects.

The unitary for a counter-propagating pulse pair, with first pulse direction $z$, can be expressed in the computational basis $\{ e, g \}$:
\begin{widetext}
\begin{align}
U_{\text{pair}} (z, \xi) = \left( 
\begin{array}{cc}
e^{-i z k x} (\cos(kx)\cos(\pi \xi) + i z \sin(kx) ) & \cos(kx) \sin(\pi \xi) (-i \cos(\phi) + \sin(\phi)) \\
  \cos(kx) \sin(\pi \xi) ( -i \cos(\phi) - \sin(\phi) )  &  e^{i z k x} ( \cos(kx) \cos(\pi \xi) - i z \sin(kx) )     
  \end{array}
  \right), \label{eq:puncomp}
\end{align}
\end{widetext}
such that $\xi=1$ gives
\begin{align} 
U_{\text{pair}} (z, 1) = - \left( 
\begin{array}{cc}
e^{-2 i z k x}  &  0  \\  0  &  e^{2 i k x}  \end{array} \right),
\end{align}
with the expected state-dependent momentum kicks and no $\phi$-dependence.
The $\phi$-dependence for imperfect pulses is in the terms of equation~(\ref{eq:puncomp}) corresponding to population transfer between internal states, and represents the angle of rotation on the Bloch sphere. 
It does not affect the magnitude of rotation which provides the error, and we set $\phi=0$ for simplicity.

The motional and internal operators commute for separate ions, and the unitary for a two-ion imperfect $\pi$-pulse is given by
\begin{multline}
U_{\text{2pulse}} (z) \\
= e^{\frac{-i \xi \pi}{2} (\sigma_1^+ e^{i z k x_1} + \sigma_1^- e^{-i z k x_1} + \sigma_2^+ e^{i z k x_2} + \sigma_2^- e^{-i z k x_2} )}.
\end{multline}
Using this unitary we construct the evolution from pulse pairs, which we intersperse with the motional free evolution unitaries to build up our gate operations.
The necessary pulse times for gates with varying numbers of pulses are found according to the applied scheme~\cite{GZC03PRL,Bent15NJP}.

\begin{figure}[t!]
     \centering
     \includegraphics[width= \columnwidth]{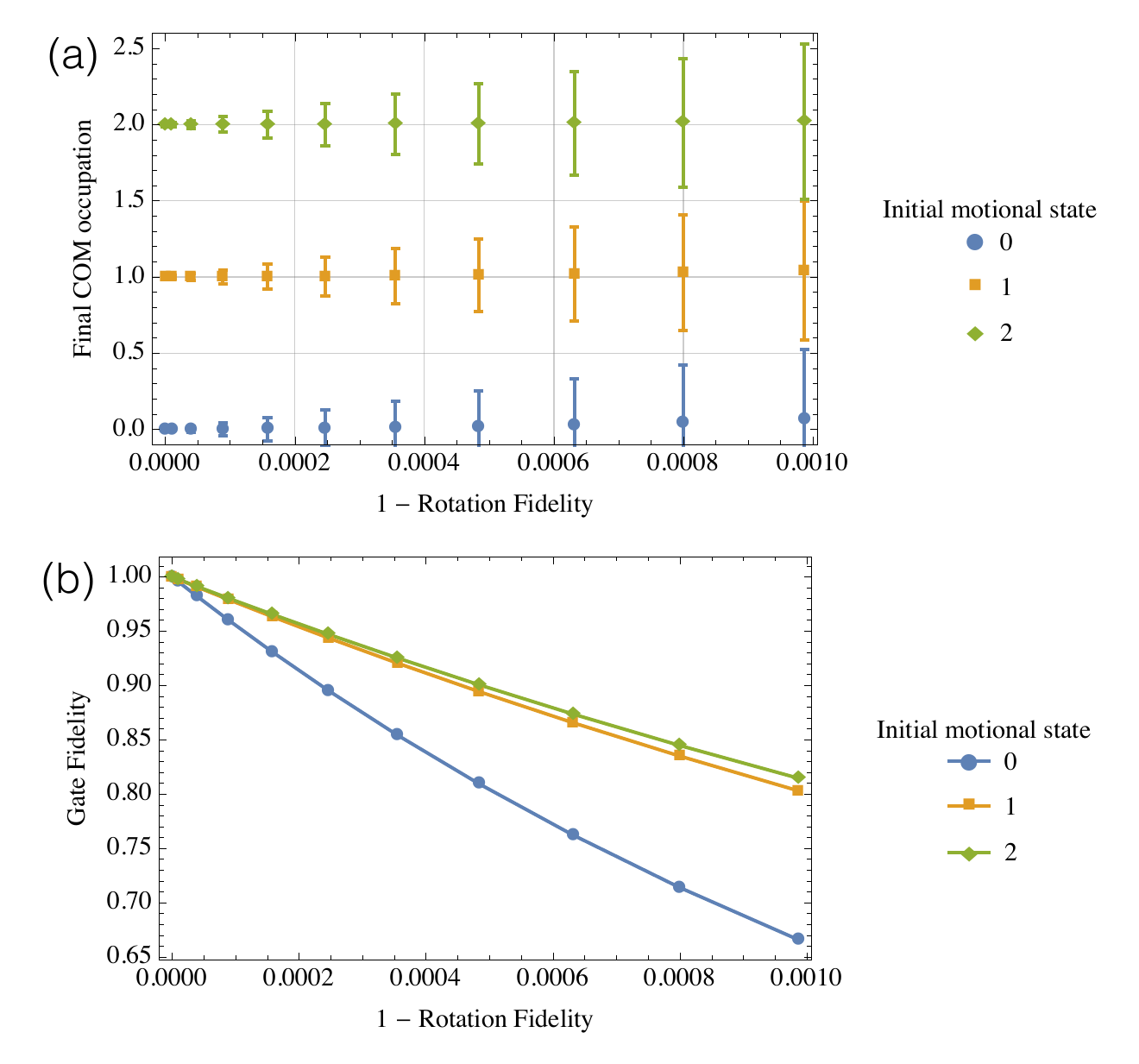}
   \caption[Fast gate infidelity and motional occupation due to imperfect pulse areas]{A GZC gate with $n=1$ is applied with varying rotation fidelity for individual pulses and different initial motional occupation.
   (a) The mean and standard deviation (error bars) in the occupation of motional states following the gate are shown.
   (b) Gate fidelity is shown as a function of pulse rotation fidelity.
   }
   \label{fig:n1pulses}
\end{figure}

Figure~\ref{fig:n1pulses} shows the effect of the initial motional state on final mode occupation and gate fidelity for a GZC gate with $n=1$.
Increasing infidelity in individual $\pi$-pulses, or rotation infidelity, damages the full gate fidelity and increases both the mean and standard deviation of the mode occupation after the gate.
The initial motional state before the gate is applied affects the magnitude of the gate infidelity.
There is not a clear relationship between initial motional state and infidelity; however each initial state is harmed by pulse errors.
Rotation infidelity around $3 \times 10^{-4}$ is required for gate fidelity better than 0.9, or rotation infidelity around $10^{-5}$ for a gate fidelity above 0.99.

\begin{figure*}[t]
     \centerline{\includegraphics[width= 2\columnwidth]{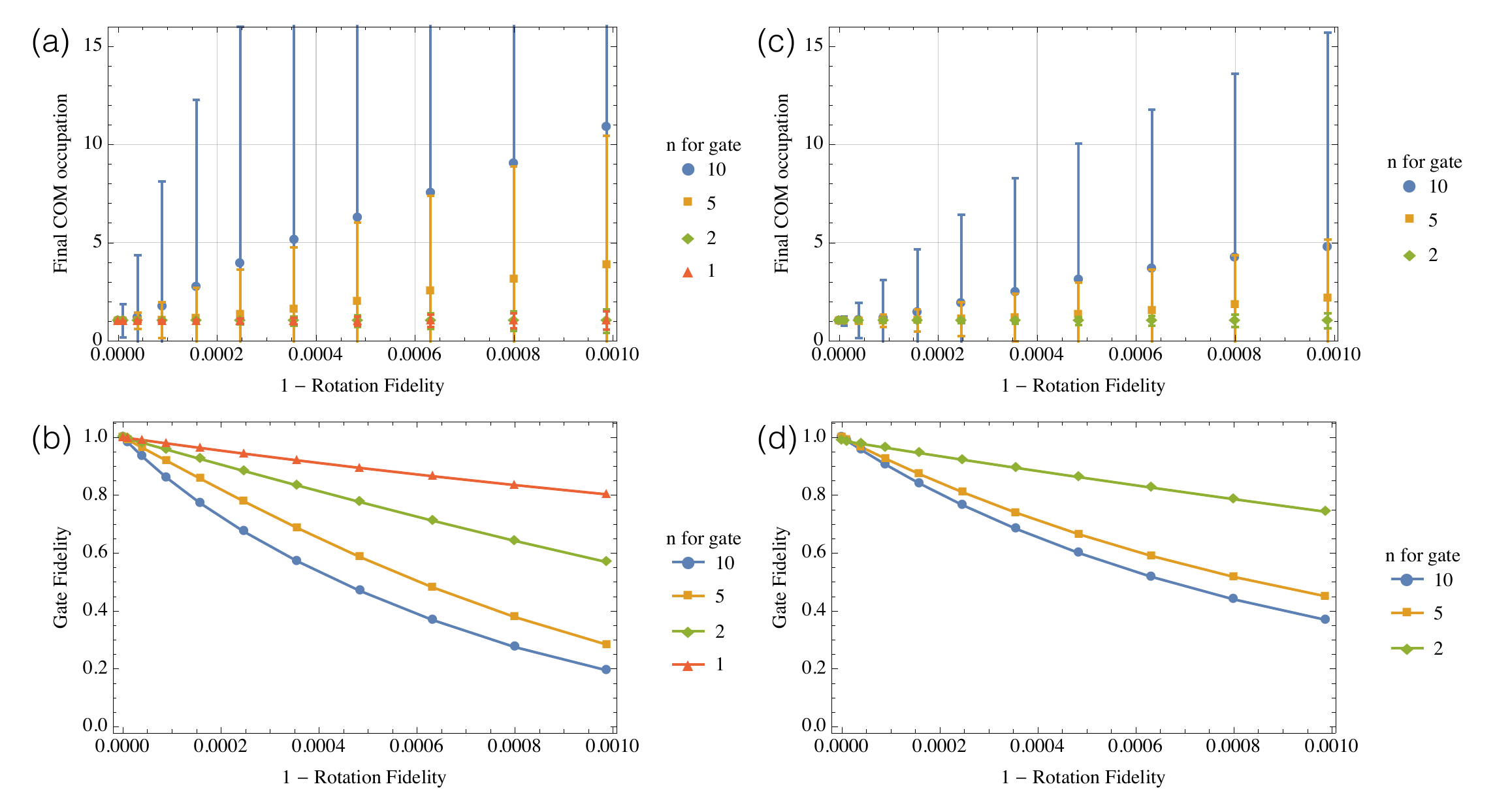}}
   \caption[Pulse errors: gate infidelity from different pulse numbers]{
   A GZC (a,b) and FRAG (c,d) fast gate are applied to $|ee\rangle |1\rangle_c |1\rangle_r$ with varying $n$ and pulse rotation error.
   (a,c): The mean and standard deviation (error bars) in the occupation of the COM mode are shown following the gate applied to the $|ee\rangle$ internal state.
   (b,d): Gate fidelity is shown as a function of pulse rotation infidelity.
   }
   \label{fig:npulses}
\end{figure*}

Higher numbers of perfect $\pi$-pulses provide faster gate times, more stability, and improved scalability.
However, as the number of pulses in the gate increases with $n$, the errors in each pulse cause compounding gate infidelities, shown in Figure~\ref{fig:npulses}.
For both the FRAG and GZC gates, Figure~\ref{fig:npulses} shows dramatic increases in the mean and standard deviation of the motional state following a gate as the number of pulses increases.
For each scheme, with $n \lesssim 10$, rotation infidelity less than $10^{-5}$ is required for a gate fidelity above 0.99, or rotation infidelity less than $10^{-4}$ is required for a gate fidelity above 0.9.
Both schemes are similarly affected by pulse error compounding with pulse number.

\begin{figure}[t]
     \centering
     \includegraphics[width= \columnwidth]{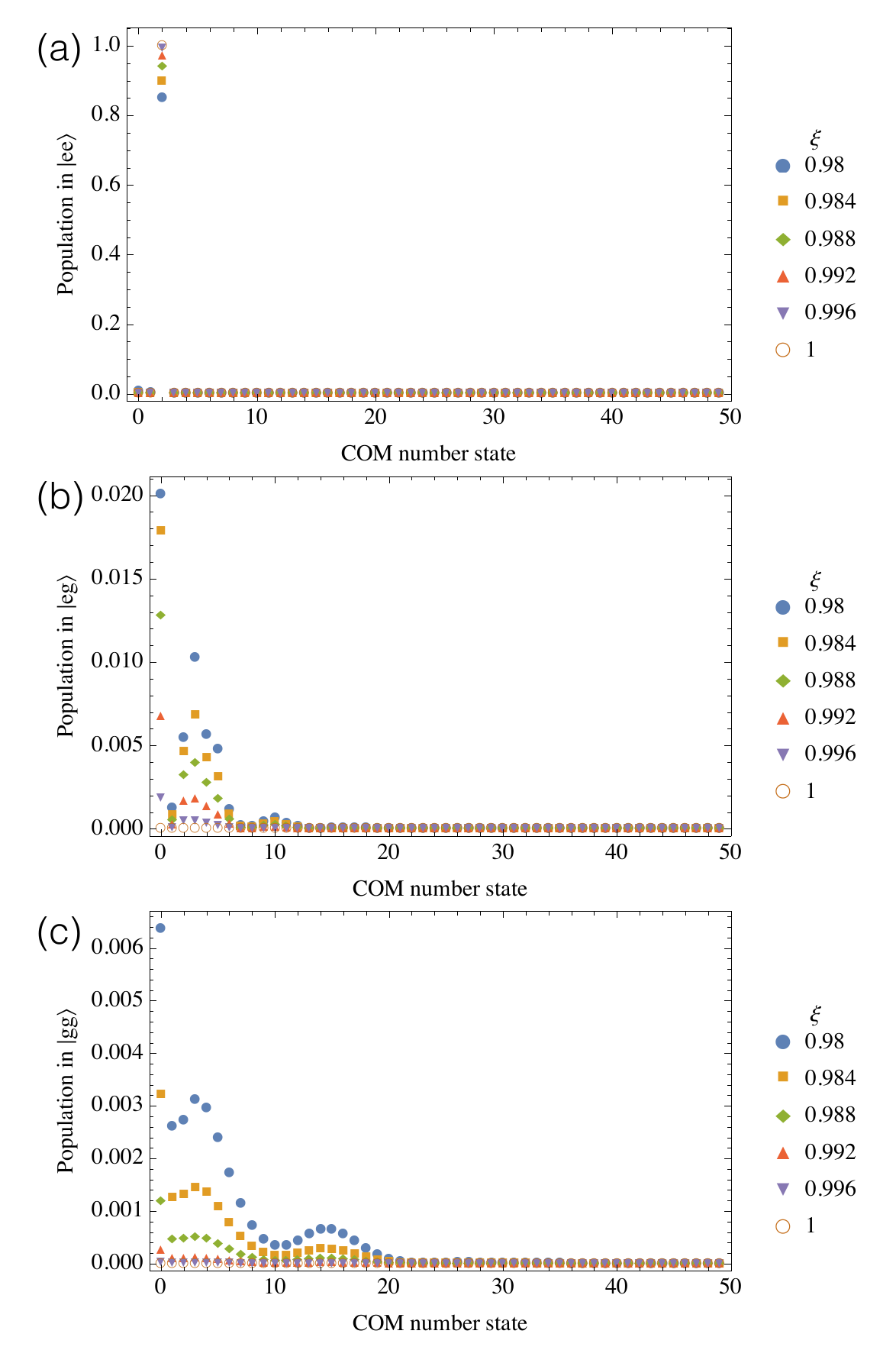}
   \caption[Population transfer from a gate composed of imperfect pulse areas]{Population in (a) $|ee\rangle$, (b) $|eg\rangle$ and (c) $|gg\rangle$ states after a GZC gate applied to $|ee\rangle \otimes |2\rangle_c$ with $n=1$.
   The fraction $\xi$ of a perfect square $\pi$-pulse performed determines the restoration of the internal state and COM motional mode to the initial state.
   }
   \label{fig:popspulse}
\end{figure}

Using square pulses, where the pulse area is proportional to $\xi$, we can find the pulse area stability requirements.  
Systematic pulse area error $(1-\xi)$ on the order of 0.4\% is permissible for fidelity better than 0.9 and $n \lesssim 10$ for each scheme.
Pulse area error $(1-\xi ) \leq 0.2\%$ is required for a fidelity above 0.98.
Figure~\ref{fig:popspulse} demonstrates the impact of systematic pulse-area errors on the internal state and mode occupation following a GZC gate with $n=1$; population is lost to other internal states with variable motional mode occupation.

\section{Conclusions} \label{sec:conc}

The duration of fast gates directly impacts their fidelity and scalability with the number of trapped ions.
We have presented a technique for calculating gate fidelities to first order in the error for large numbers of applied pulses.
Applying this technique to two ions, we have demonstrated that pulse errors cause compounding infidelity with the number of pulses composing the gate.
Gate duration scales with the number of pulses, so this pulse fidelity requirement is of great importance for using fast gates for scalable QIP.
Pulse infidelity less than $10^{-5}$ is required for gate fidelity above 0.99 with up to 140 pulse pairs in the FRAG and GZC gate schemes.
We have also shown that different numbers of applied pulses do not significantly alter the valid RWA regime: pulse durations much longer than the atomic transition period are required.
Experimental implementation of a fast gate, which requires fast and robust $\pi$-pulses, will be a significant step towards large-scale QIP with ions.

\section{Acknowledgments}
This work was supported by the Australian Research Council Centre of Excellence for Quantum Computation and Communication Technology (Project number CE110001027) (ARRC), Australian Research Council Future Fellowship (FT120100291) (JJH) as well as DP130101613 (JJH, ARRC).

\bibliographystyle{ieeetr} 
\bibliography{errorlib}

\end{document}